\documentstyle[12pt]{article}
\begin{document}
\setlength{\baselineskip}{0.8cm}
\newcommand{\beq}{\begin{equation}}
\newcommand{\eeq}{\end{equation}}
\newcommand{\beqn}{\begin{eqnarray}}
\newcommand{\eeqn}{\end{eqnarray}}
\newcommand{\beqnn}{\begin{eqnarray*}}
\newcommand{\eeqnn}{\end{eqnarray*}}
\def\alf{\alpha}
\def\alfp{\alpha \prime}
\def\argt{\frac{t - t_0}{\tau}}
\def\lb#1{\label{#1}}
\def\3cdot{\cdot \cdot \cdot}
\def\pd#1#2{\frac{\partial #1}{\partial #2}}
\def\td#1#2{\frac{d #1}{d #2}}
\def\dtot#1{\frac{d #1}{dt}}
\def\seq{Schr\"odinger equation}
\def\g{\gamma}
\def\G{\Gamma}
\def\d#1{\delta_{#1}}
\def\half{\frac{1}{2}}
\def\l{\lambda}
\def\L{\Lambda}
\def\z{\zeta}
\def\om0{\omega _0}
\def\Om0{\Omega _0}
\def\ket#1{|#1 \>}
\def\bra#1{\< #1|}
\def\x{\times}
\def\>{\rangle}
\def\<{\langle}
\def\text#1{{\rm{#1}}}
\def\sig{\sigma}
\def\Sig{\Sigma}
\def\o{\omega}
\def\O{\Omega}
\def\d{\delta}
\def\D{\Delta}
\def\t{\theta}
\def\T{\Theta}
\def\del{\partial}
\def\->{\rightarrow}
\def\=>{\Rightarrow}
\def\-->{\longrightarrow}
\def\==>{\Longrightarrow}
\def\sech{\mbox{sech}}
\def\veps{\varepsilon}
\def\eps{\epsilon}
\def\mcal#1{\mathcal{#1}}
\def\rpar{\right)}
\def\lpar{\left(}
\def\lbk{\left[}
\def\rbk{\right]}
\def\lbr{\left\{}
\def\rbr{\right\}}
\def\dag{\dagger}
\def\bb#1{\mbox{\boldmath{$#1$}}}
\def\bbR{\mbox{\boldmath{$R$}}}
\def\bbrho{\mbox{\boldmath{$\rho$}}}
\def\bbF{\mbox{\boldmath{$F$}}}
\def\bbS{\mbox{\boldmath{$S$}}}
\def\bbP{\mbox{\boldmath{$P$}}}
\def\bbQ{\mbox{\boldmath{$Q$}}}
\def\bbx{\mbox{\boldmath{$x$}}}
\def\bbX{\mbox{\boldmath{$X$}}}
\def\bba{\mbox{\boldmath{$a$}}}
\def\bbT{\mbox{\boldmath{$T$}}}
\def\bbZ{\mbox{\boldmath{$Z$}}}
\def\bbD{\mbox{\boldmath{$D$}}}
\def\til{\tilde}
\def\rf#1{(\ref{#1})}
\def\nn{\nonumber}
\title{Finite-length soliton solutions of the local homogeneous nonlinear
 Schr\"odinger equation}
\author{E.C. Caparelli\thanks{E-mail capareli@ifsc.sc.usp.br},
        V.V. Dodonov\thanks{E-mail vdodonov@power.ufscar.br} and
        S.S. Mizrahi\thanks{E-mail salomon@power.ufscar.br} \\
{\sl Departamento de F{\'{\i}}sica, CCT, Universidade Federal de
        S\~ao Carlos,}\\
{\sl Via Washington Lu{\'{\i}}s km 235, C.P.676, CEP.13.565-905}}
\maketitle
\begin{abstract}
We found a new kind of soliton solutions for the 5-parameter family of
the potential-free
Stenflo-Sabatier-Doebner-Goldin nonlinear modifications of the
Schr\"odinger equation.
In contradistinction to the ``usual'' solitons like
$\left\{\cosh\left[\beta(x-kt)\right]\right\}^{-\alpha}\exp[i(kx-\o t)]$,
the new {\em Finite-Length Solitons} (FLS) are nonanalytical
functions with continuous first derivatives, which
are different from zero only inside some finite regions of space.
The simplest one-dimensional example is the function which is equal to
$\left\{\cos\left[\gamma(x-kt)\right]\right\}^{1+\d}\exp[i(kx-\o t)]$
(with $\d>0$) for $|x-kt|<\pi/2\gamma$, being identically equal to zero
for $|x-kt|\ge \pi/2\gamma$. The FLS exist even in the case of a weak
nonlinearity, whereas the ``usual'' solitons exist provided the
nonlinearity parameters surpass some critical values.
\end{abstract}

\vskip 1.5cm

PACS Ref: 03.65.-w, 52.35.Sb, 02.30.Jr

\newpage
\section{Introduction}

Recently, different authors
\cite{sten82,sten86,sten88,sten89,sten91,sab90,sab91,sab94,dg92,g92,dg94}
discovered an interesting multiparametric nonlinear
homogeneous modification of the Schr\"odinger equation in the coordinate
representation.
In the most general form this
{\it Stenflo--Sabatier--Doebner-Goldin\/} (SSDG) equation
can be written as
 (we confine ourselves to the case of a free motion and assume $\hbar=m=1$)
\beq
i \frac{\partial\psi}{\partial t}=-\frac{1}{2}\nabla^{2}\psi+
\Omega\{\psi\}\psi \lb{1}
\eeq
where the local nonlinear functional $\Omega\{\psi\}$
is as a linear combination of terms
$\Delta\psi/\psi$, $(\nabla\psi/\psi)^2$, $ |\nabla\psi/\psi|^2$
and their complex conjugated counterparts, so that $\Omega\{\psi\}$
satisfies the {\it homogeneity condition\/}
 $\Omega\{\gamma\psi\}
=\Omega\{\psi\}$ for an arbitrary complex constant $\gamma$.
The specific choices of the
{\it complex\/} coefficients in the linear combination correspond to the
equations describing waves in plasmas with sharp boundaries and in
nonlinear media \cite{sten82,sten86,sten88,sten89,sten91}.
However, trying to interpret \rf{1} as a {\it quantum mechanical\/}
equation one must worry about the conservation of probability.
For this reason, the functional $\Omega\{\psi\}$ was chosen in
\cite{sab90,sab91,sab94} in an explicit real form:
$\Omega\{\psi\}= \hat{\cal D} \ln|\psi|$,
where $\hat{\cal D}$ is the second order differential operator
$
\hat{\cal D}f = a\Delta f +{\bf b}\cdot\nabla f +c\nabla f\cdot\nabla f
$,
with real parameters $a,{\bf b},c$.
However, it was shown in \cite{dg92,g92} that the normalization could be
saved even in the presence of
{\it imaginary\/} (antihermitian) nonlinear corrections of a special
kind. The most general parametrization was proposed in \cite{dg94},
where $\Omega\{\psi\}$ was written in terms of real and imaginary parts,
$\Omega\{\psi\}=R\{\psi\}+iI\{\psi\}$, as follows,
\beq
I\{\psi\}=\frac{1}{2} D\frac{\nabla^{2}(\psi^*\psi)}{\psi^*\psi}\,,
\lb{imag}
\eeq
\beq
R\{\psi\}= \tilde D\sum_{j=1}^5 \l_j\L_j[\psi]=
\tilde D\sum_{j=1}^5 c_j R_j[\psi]\,,
\lb{paramdg}
\lb{real}
\eeq
where all the coefficients $\l_j$ and $c_j$ are real, and the functionals
$\L_j[\psi]$ or $R_j[\psi]$ are expressed in terms of the derivatives of the
wave function or in terms of the probability density $\rho=\psi^*\psi$ and
the probability current
${\bf j}=\left(\psi^*\nabla\psi-\psi\nabla\psi^*\right)/2i$:

\begin{tabular}{ll}
$\Lambda_{1}[\psi]=\displaystyle{\mbox{Re}\left(\frac{\nabla^{2}\psi}{\psi}
\right)}$&\quad
$R_1[\psi] =\displaystyle{\frac{\nabla\cdot{\bf j}}{\rho}}$\\[3mm]
$\Lambda_{2}[\psi]=\displaystyle{\mbox{Im}\left(\frac{\nabla^{2}\psi}{\psi}
\right)}$&\quad
$R_2[\psi] =\displaystyle{\frac{\nabla^2\rho}{\rho}}$\\[3mm]
$\Lambda_{3}[\psi]=\displaystyle{\mbox{Re}\left(\frac{\nabla\psi}{\psi}
\right)^{2}}$&\quad
$R_3[\psi] =\displaystyle{\frac{{\bf j}^2}{\rho^2}}$\\[3mm]
$\Lambda_{4}[\psi]=\displaystyle{\mbox{Im}\left(\frac{\nabla\psi}{\psi}
\right)^{2}}$&\quad
$R_4[\psi] =\displaystyle{\frac{{\bf j}\cdot\nabla\rho}{\rho^2}}$\\[3mm]
$\Lambda_{5}[\psi]=\displaystyle{\frac{|\nabla\psi|^{2}}{|\psi|^{2}}}$&\quad
$R_5[\psi] =\displaystyle{\frac{(\nabla\rho)^2}{\rho^2}}$\\[3mm]
\end{tabular}

\noindent
The coefficients $\l_j$ and $c_j$ are related as follows,

\vspace{3mm}

\begin{tabular}{ll}
$\l_1=2c_2$ &\quad $c_1=\l_2$\\
$\l_2=c_1$ &\quad $c_2=\frac12\l_1$\\
$\l_3=2c_5-\frac12 c_3$ &\quad $c_3=\l_5-\l_1-\l_3$\\
$\l_4=c_4$ &\quad $c_4=\l_4$\\
$\l_5=2c_2+2c_5+\frac12 c_3$ &\quad $c_5=\frac14(\l_5+\l_3-\l_1)$
\end{tabular}

\vspace{3mm}
\noindent
More general homogeneous nonlinear functionals, which include as special cases
the nonlocal terms proposed by Gisin \cite{gis} and by
Weinberg \cite{wein}, were given in \cite{dm93,dm95,grig,dm98}.

It is not clear, until now, whether nonlinear corrections to the
Schr\"odinger equation of the SSDG type have a physical meaning from the
point of view of quantum mechanics (possible experiments which could
verify the existence of such corrections were proposed in \cite{dmpla,dmnew},
and the relations between the SSDG-equation and the master equation for
mixed quantum states were studied in \cite{dm95,dmnew,dmcla}).
Nonetheless, the mathematical structure of the new family of nonlinear
equations appears rather rich. In particular, studying this
family resulted recently in discovering the nonlinear gauge transformations
\cite{g95,dg96,g97}.

The aim of our article is to show another remarkable property of the SSDG
equation, namely, the existence of a {\it new type of soliton solutions\/},
which are different from zero {\it in a finite space domain\/} even for
{\it arbitrarily small\/} nonlinear coefficients. As far as we know, such
kind of solitons was not discussed earlier.
%
\section{Soliton solutions with linear phase}
%
Looking for a {\it shape invariant\/} solution to the SSDG equation
\rf{1}--\rf{real} with a {\it linear phase\/},
\beq
\psi({\bf x},t)=g({\bf x}-{\bf v}t)e^{i({\bf k}{\bf x}-\o t)},
\lb{sol1}
\eeq
we obtain the following two equations for the {\it real\/} function
$g({\bf x})$:
\beq
({\bf k}-{\bf v})\frac{\nabla g}{g} =D\left[\frac{\nabla^2 g}{g} +
\left(\frac{\nabla g}{g}\right)^2\right]
\lb{eqim}
\eeq
\beq
(1-\sigma) \frac{\nabla^2 g}{g} -\xi \left(\frac{\nabla g}{g}\right)^2
-2\mu {\bf k} \cdot \frac{\nabla g}{g}
={\bf k}^2(1+\eta) -2\o \,,
\lb{eqre}
\eeq
where the new coefficients are defined as
\[
\sigma=2\tilde{D}\l_1 \equiv 4\tilde{D}c_2,
\quad \xi=2\tilde{D}\left(\l_3+\l_5\right)
\equiv 4\tilde{D}\left(c_2+2c_5\right),
\]
\[
\eta =2\tilde{D}\left(\l_5 -\l_3-\l_1\right)\equiv 2\tilde{D}c_3\,,
\quad \mu= 2\tilde{D}\left(\l_2+\l_4\right)
\equiv 2\tilde{D}\left(c_1+c_4\right).
\]

A general solution to eq. \rf{eqim} in the one-dimensional case is
\[
g_D(x)=\left\{C_1 + C_2 \exp[(k-v)x/D]\right\}^{1/2},
\]
with arbitrary constants $C_1$ and $C_2$. However, the function
$g_D(x)$ cannot be normalized, thus in order to guarantee normalization we
impose  
\[
{\bf k}={\bf v}, \quad D=0,
\]
i.e. soliton solutions can only exist in the absence of dissipative terms
in the Hamiltonian.

The substitution
\beq
g({\bf x})=[f({\bf x})]^{\alpha}, \quad
\alpha=\frac{1-\sigma}{1-\sigma-\xi} \, ,
\lb{alfa}
\eeq
eliminates the nonlinear term $(\nabla g/g)^2$ in eq. \rf{eqre}, such that
\beq
\nabla^2 f -2\mbox{\boldmath{$\kappa$}}\cdot \nabla f +\gamma^2 f=0 \; ,
\lb{lineq}
\eeq
where
\beq
\gamma^2 =\left[ 2\o -{\bf k}^2(1+\eta)\right]
\frac{1-\sigma-\xi}{(1-\sigma)^2}\,,
\quad \mbox{\boldmath{$\kappa$}}=\frac{2\mu{\bf k}}{1-\sigma}\,.
\lb{defgam}
\eeq
Note that $\gamma^2$ is a free parameter, which may assume both positive and
negative values, depending on the packet average energy
\[
\langle E\rangle \equiv i\int_{-\infty}^{\infty}\psi^{*}({\bf x},t)
\frac{\partial \psi({\bf x},t)}{\partial t}\,d{\bf x} =\o.
\]

A general solution to eq. \rf{lineq} in the one-dimensional case reads
\beq
f(x)=e^{\kappa x}\left(C_1 e^{sx} +C_2 e^{-sx}\right),
\quad s=\sqrt{\kappa^2-\gamma^2}.
\lb{sols}
\eeq
In particular,

\noindent ({\em a})
For $\gamma^2<0$, function \rf{sols} goes to infinity when
$x\to\pm\infty$ (if both constants $C_1$ and $C_2$ are positive),
so a normalizable solution $g(x)$ (eq. \rf{alfa}) exists only under the
condition $\alpha<0$,
i.e., for parameters $\sigma$ and $\xi$ satisfying the inequalities
$\sigma<1,\; \xi>1-\sigma$ or
$\sigma>1,\; \xi<1-\sigma$ (in other words, these parameters must be
located between the straight lines
$\sigma=1$ and $\sigma+\xi=1$ in the $\sigma \xi$-plane).
This means that only
{\it strong nonlinearity\/} can give ``usual'' soliton solutions with
exponentially decreasing tails, whose simplest representative ($\mu =0$)
reads
\beq
g_*(x)= \left[ \cosh(\beta x)\right]^{-|\alpha|}.
\lb{sol2}
\eeq
This conclusion agrees with the results of studies \cite{sab90,sab91,sab94},
where exponentially confined solitons were found for the nonlinear
functionals like $\Omega\{\psi\}= a\Delta (\ln|\psi|)$.
Similar solutions to the special cases of the
SSDG equation with complex coefficients were studied in
\cite{sten88,sten89,sten91}.
A large family of exact solutions corresponding to the most general
Doebner--Goldin parametrization \rf{imag}-\rf{real} was found in
\cite{natt94,natt95,natt96}. However, that family
does not contain the solitons with a linear phase. For example,
the solution given in \cite{natt95} has the same amplitude factor as in
\rf{sol2}, but its phase is proportional to $\ln[g(x-kt)]$, so it does
not converge to the plane-wave solution of the linear Schr\" odinger
equation when the nonlinear coefficients $D$ and $\til{D}$ go to zero.

\noindent ({\em b}) For $ 0 < \gamma^2<\kappa^2$, function
\rf{sols} goes to infinity only for $x\to+\infty$, while for $x\to-\infty$
it goes to zero (or vise versa). In this case, we cannot obtain
a normalizable solution in the form \rf{alfa}
for any value of $\alpha$.

\noindent ({\em c}) Quite different situation arises when
$\gamma^2>\kappa^2$, then expression \rf{sols} shall contain
{\it trigonometric\/} functions, and (making a
shift of the origin, if necessary) we arrive at a solution to eq.
\rf{eqre} in the form
\beq
g_{\d}(x)=\left[e^{\kappa x}\cos(\til\gamma x)\right]^{1+\d}
\lb{FLS}
\eeq
with $\til\gamma=\sqrt{\gamma^2-\kappa^2}\ge 0$ and
\beq
\d=\frac{\xi}{1-\sigma-\xi}\,.
\lb{defdelt}
\eeq
At first glance, we have a problem when $f<0$, since function $f^{\alpha}$
is ill-defined in this case (unless the exponent $\alpha$ is an integer).
But we notice that if $\alpha>1$ (i.e. $\d>0)$, then the function
$g(x)=[f(x)]^{\alpha}$ turns into zero
{\it together with its derivative\/} $ g'(x)$ at $\til{\g} x = \pi /2$.
 This means that there exists an
integrable solution with a {\it continuous first derivative\/},
which is {\it localized completely\/} inside a {\it finite\/} domain:
\[
\psi_{\d}(x,t)= \left\{ \begin{array}{cl}
\left\{\cos\left[\til\gamma(x-kt)\right]\right\}^{1+\d}
\exp\left[(1+\d)\kappa(x-kt) + i(kx-\o t)\right]
&\mbox{if} \quad |\til\gamma(x-kt)| < \pi/2\\[3mm]
0 & \mbox{if} \quad  |\til\gamma(x-kt)|\ge \pi/2
\end{array} \right.
\]
It is remarkable that such a ``finite-length soliton'' (FLS) exists for an
{\it arbitrarily weak\/} nonlinearity, since the requirement $\d>0$ implies
the inequalities
\beq
0<\xi < 1-\sigma
\lb{conds}
\eeq
which can be satisfied for small values of $\xi$ and $\sigma$
(another possibility is $0>\xi > 1-\sigma$, but it demands $\sigma>1$,
meaning a stronger nonlinearity).
In terms of the coefficients $\l_j$ and $c_j$ , condition \rf{conds} reads
\beqnn
& \til{D}\left(\l_3+\l_5\right)>0,&
\quad 2\til{D}\left(\l_1+\l_3+\l_5\right)<1\\
& \til{D}\left(c_2+2c_5\right)>0, & \quad 8\til{D}\left(c_2+c_5\right)<1 \; .
\eeqnn
It was shown in \cite{dg94} that the SSDG equation is Galilean invariant
provided that (i) $c_1+c_4=0$ and (ii) $c_3=0$. In our notation this
means $\mu=\eta=0$. Thus we arrive at the 3-parameter family of homogeneous
local nonlinear functionals admitting Galilean-invariant and spatially
confined soliton solutions to eq. \rf{1}:
\beqn
\O\{\psi\}&=&\frac12\left\{
\sigma\,\mbox{Re}\frac{\nabla^{2}\psi}{\psi}+
\nu\,\mbox{Im}\left[\nabla\cdot\left(\frac{\nabla\psi}{\psi}\right)\right]+
\xi\left[\mbox{Re}\frac{\nabla\psi}{\psi}\right]^2 +
\sigma\,\left[\mbox{Im}\frac{\nabla\psi}{\psi}\right]^2
\right\}
\lb{finlam}\\
&=&\frac18\left\{
2\sigma\,\frac{\nabla^{2}\rho}{\rho}+
4\nu\,\nabla\cdot\left(\frac{{\bf j}}{\rho}\right)+
(\xi-\sigma)\left(\frac{\nabla\rho}{\rho}\right)^{2}
\right\}.
\lb{finc}
\eeqn
Note that parameter $\nu=2\tilde{D}\l_2=2\tilde{D}c_1$ does not make any
influence on the discussed solutions, thus only derivatives of the density
$\rho$ and not of the current density ${\bf j}$ are important for soliton
solutions. The only crucial parameter is
$\xi$, so, the simplest 1-parameter nonlinear functional admitting
FLS solution reads ($\nu = \sigma =0$),
\beq
\O\{\psi\}=\frac{\xi}{2}\left[\mbox{Re}
\left(\frac{\nabla\psi}{\psi}\right)\right]^{2}=
\frac{\xi}{8}\left(\frac{\nabla\rho}{\rho}\right)^{2}, \quad 0<\xi<1\, ,
\lb{simpl}
\eeq
The explicit form of all FLS-solutions is as follows,
\beq
\psi_{{\bf k}\gamma}({\bf x},t)=\left\{
\begin{array}{cl}
\left[f_{\gamma}({\bf x}-{\bf k}t)\right]^{1+\d}
e^{i({\bf k}{\bf x}-\o_{{\bf k}\gamma} t)} &
\mbox{if} \quad |{\bf x}-{\bf k}t|\in {\cal R}^{(+)}(f_{\gamma})\\[3mm]
 0 & \mbox{if} \quad |{\bf x}-{\bf k}t|\notin {\cal R}^{(+)}(f_{\gamma})
\end{array}
\right.,
\lb{solFLS}
\eeq
where $f_{\gamma}({\bf x})$ is any {\it positive\/} solution to the
Helmholtz equation $\left(\nabla^2 +\gamma^2\right)f=0$ with an arbitrary
real constant $\gamma$, and ${\cal R}^{(+)}(f_{\gamma})$ is the internal
part of a space region bounded by a closed surface (in 3 dimensions) or a
closed curve (in 2 dimensions) determined by the equation
$f_{\gamma}({\bf x})=0$ (in principle, this region may be multi-connected).
To avoid any ambiguity, we define the nonlinear functional $\O\{\psi\}$
for $\psi=0$, assuming $\O\{\psi\}\psi=0$ on such points.
Although the solution \rf{solFLS} is non-analytical, it has continuous first
derivatives in all points of the space.

Under the Galilean invariance symmetry ($\eta=0$), the frequency
$\o_{{\bf k}\gamma} $ (eq. \rf{defgam}) equals
\beq
\o_{{\bf k}\gamma}=\frac12{\bf k}^2 +
\frac12\gamma^2\frac{(1-\sigma)^2}{1-\sigma-\xi}\,, \lb{galinvfreq}
\eeq
so, the usual dispersion relation of linear Quantum Mechanics
($\o_{{\bf k}}=\frac12 {\bf k}^2$) is modified
by an additional constant term (proportional to $\gamma^2$) that may be
interpreted as an ``internal energy'' of the wave packet \rf{solFLS}
due to its
confinement, whereas ${\bf k}^2/2$ is the energy of the ``center-of-mass''
motion. For $\gamma=0$, then $\nabla^2 f_0 = 0$, and considering $f_{0}=1$
we obtain a plane wave solution to the Schr\"odinger equation
with the wave number ${\bf k}$.

The concrete shapes of the FLS-packets in 2 and 3 dimensions may be quite
diverse. The most symmetric solutions are given by eq. \rf{solFLS} with
$f_{\gamma}({\bf x})$ in the form of
$J_0(\gamma |{\bf x}|)$ or $\sin(\gamma|{\bf x}|)/|{\bf x}|$
(in 2 and 3 dimensions, respectively). However, there exist also asymmetric
packets with functions $f_{\gamma}({\bf x})$ proportional to
$J_m(\gamma |{\bf x}|)\cos(m\varphi)$ or
$j_l(\gamma |{\bf x}|)Y_{lm}(\vartheta,\varphi)$, where $J_m(x)$ is the
Bessel function, $j_l(x)$ is the spherical Bessel function (proportional to
the Bessel function with the semi-integral index), and
$Y_{lm}(\vartheta,\varphi)$ is a real-valued analog of the spherical harmonics,
$\vartheta,\varphi$ being the usual angular variables.
%
\section{Discussion}
%

Although the substitution \rf{alfa} was used already in \cite{sab94,dg94},
the existence of the FLS solutions was not noticed before, perhaps, because
the authors of the cited papers were looking for solutions of the
stationary Schr\"odinger equation or for usual exponentially confined
solitons.
It should be noted that the substitution \rf{alfa} linearizes only the
equation \rf{eqre} for the {\it real\/} amplitude of the special form of
solution \rf{sol1}, but not eq. \rf{1} as a whole.
As was shown in \cite{g95,dg96}, the {\it nonlinear gauge transformation\/}
(NGT)
\beq
\psi \mapsto \psi' =|\psi|\exp\lbk i\lpar z^* \ln\psi+ z \ln\psi^*
 \rpar \rbk
\lb{gauge}
\eeq
(where $z$ is a complex parameter) transforms any SSDG equation into an
equation of the same kind, but with another set of coefficients, and all
equations can be classified in accordance with the possible values of 5
invariants of the NGT family. In the case of eq. \rf{1} with the
functional \rf{simpl} the invariants are as follows (we use the same notation
as in \cite{dg96}):
\[
\tau_1=\tau_4=0, \quad \tau_2=\frac18, \quad \tau_3=-1, \quad
\imath_5=-\,\frac{\xi}{16}
\,.
\]
If we had $\imath_5=0$, then the whole SSDG equation could be linearized
by means of a
suitable NGT. It is the nonzero value of parameter $\xi$ that prevents the
linearization and makes possible the existence of the FLS solutions.
It is interesting to note in this connection, that various special cases
of the general SSDG equation were considered before the general structure
of the equation was found in \cite{sab94,dg94}, but the coefficients
were chosen in such a way that the parameter $\xi$ was almost always taken
equal to zero
\cite{guer,smol,vig}. The only exception is Ref. \cite{kib}, where the
only nonzero coefficient in the $\l_j$-parametrization is $\l_5$; however,
in this case not only $\xi\neq 0$, but $\eta\neq 0$, too, so the
Galilean invariance symmetry is absent. 

Turning to a possible physical meaning of the FLS solutions, we may say that
they realize the ``dream of De Broglie'', in a sense that they permit
to identify a quantum particle with a nonspreading wave packet of finite
length travelling with a constant velocity in the free space.
Earlier, the only proposed nonlinear equation that resulted into a
nonspreading wave packet solution for
a free particle was the one proposed by Bialynicki-Birula and
Mycielski \cite{bbm} (BBM), with the nonlinear term
$\O\{\psi\}=-b\ln|\psi|^2$ (it was shown recently that this term can
arise if one applies to the SSDG equation a nonlinear gauge transformation
with {\it time-dependent coefficients\/} \cite{dg96}).
The solitons of the BBM-equation are Gaussian wave packets ({\em gaussons})
whose constant width is inversly proportional to the nonlinear coefficient
$b$. In contrast to the {\em gaussons}, the FLS of the SSDG equation have
the width $\gamma^{-1}$ as a free parameter, independent of the nonlinear
coefficients. The most attractive feature of FLS solutions is that they
exist for an arbitrarily weak nonlinearity. Consequently, the
superposition principle of quantum mechanics, which is verified, from the
experimental point of view, with a limited accuracy, does not rule out the
nonlinear terms like \rf{simpl} immediately. On the contrary, new
experiments on the verification of (non?)linearity of quantum mechanics
could be proposed, which would take into account the FLS phenomenon.

\subsection*{Acknowledgements}

This research was supported by FAPESP (Brazil) project 96/05437-0.
S.S.Mizrahi thanks CNPq and FINEP, Brasil, for partial financial support.

\end{document}